\documentclass[12pt,a4paper,notitlepage]{article}

\usepackage{amssymb,amsmath}
\usepackage[dvips]{graphicx}

\newcommand{\bsl}{\boldsymbol}
\newcommand{\rr}{\mathbf{r}}

\begin{document}

\author{Yu.A.Simonov\thanks{e-mail: simonov@itep.ru}~ and
M.A.Trusov\thanks{e-mail: trusov@itep.ru} \\[4mm]
\textit{Institute of Theoretical and Experimental Physics, Moscow,
Russia}}

\title{Nucleon matrix elements and baryon masses in the Dirac orbital model}

\date{}

\maketitle

\begin{abstract}
\noindent Using the expansion of the baryon wave function in a
series of products of single quark bispinors (Dirac orbitals), the
nonsinglet axial and tensor charges of a nucleon are calculated.
The leading term yields $g_A = 1.27$ in good agreement with
experiment. Calculation is essentially parameter-free and depends
only on the strong coupling constant value $\alpha_s$.  The
importance of lower Dirac bispinor component, yielding 18\% to the
wave function normalization is stressed. As a check, the baryon
decuplet masses in the formalism of this model are also computed
using standard values of the string tension $\sigma$ and the
strange quark mass $m_s$; the results being in a good agreement
with experiment.
\end{abstract}

\newpage

Axial and tensor charges of a nucleon are important to
characterize the basic structure of the nucleon as  composed of
strongly coupled quarks \cite{Lee}--\cite{Bass}. Whereas many
baryon charateristics can be reasonably obtained in relativized
quark models where all relativistic effects are treated via
Salpeter equation \cite{Isgur} and spin corrections, the axial and
tensor charges are sensitive to the Dirac structure of quark wave
functions, and in particular, as will be shown below, to the
negative energy components. The most systematic treatment of
relativistic baryon system can be done in the three-body
Bethe--Salpeter formalism, however in this approach a rigorous
derivation ends up in the system of more than 20 integral
equations, and therefore, a drastic simplification is needed to
make realistic calculations. Recently this kind of approach was
developed in \cite{Bonn}, where quark-quark interactions have been
properly parametrised. However the relativistic objects like
negative energy component admixture $\eta$ (see below) are very
sensitive to the form of interactions, and in \cite{Bonn} $\eta$
has come out rather small. In our paper we choose another and much
simpler approach, which allows to treat all Dirac components
properly including lower ones, and to work out the resulting
$\eta$ (and axial and tensor charges) in a transparent way.

First of all, let us recall some formulas of the spin relativistic
theory. Consider a polarized free fermion with spin $\frac{1}{2}$.
The fermion polarization could be described by a spacelike
pseudovector $a^\mu$, which reduces in the fermion rest frame to
the unit 3-vector along which the fermion spin projection is equal
to $+1/2$. The pseudovector $a^\mu$ obviously meets conditions:
$a\cdot a=-1$, $a\cdot p=0$, where $p$ is the the fermion
4-momentum. The corresponding Michel--Wighimann fermion density
matrix is
\begin{equation}
\rho=\frac{\gamma p+m}{4m}\left[1-\gamma^5(\gamma a)\right]
\label{dens_mat}
\end{equation}
where $m$ is the fermion mass. Note, that here and below we
propose the plane wave amplitude being normalized to unity: $\bar
uu=1$. Using the density matrix (\ref{dens_mat}) the pseudovector
$a^\mu$ could be also defined in an arbitrary frame as the
averaged value of the Pauli--Lubansky operator:
\begin{equation}
\langle \hat W^\mu \rangle=\frac{1}{2}a^\mu, \quad \hat
W^\mu=-\frac{1}{2m}\varepsilon^{\mu\nu\rho\sigma}\hat J_{\nu\rho}
\hat P_{\sigma},\quad \varepsilon^{0123}=+1.
\end{equation}
where $\hat J_{\mu\nu}$ is the angular momentum operator.

Axial and tensor current operators are defined as follows:
\begin{equation}
\hat j_\mu^A=\hat{\bar\psi}\gamma_\mu\gamma^5\hat\psi,\quad \hat
j_{\mu\nu}^T=\hat{\bar\psi}\sigma_{\mu\nu}\gamma^5\hat\psi
\end{equation}
where $\sigma_{\mu}=\frac{1}{2}[\gamma_\mu,\gamma_\nu]$ (both
currents are hermitean). It it easy to check\footnote{The simplest
way to derive these relations is to work in the fermion rest frame
and to choose the $Oz$ axis direction being parallel to the space
part of the pseudovector $a$} that for a plane wave
\begin{equation} \label{ferm_av}
\left\langle \hat j_\mu^A \right\rangle=\bar u\gamma_\mu\gamma^5
u=-a_\mu,\quad \left\langle \hat j_{\mu\nu}^T \right\rangle=\bar u
\sigma_{\mu\nu}\gamma^5 u=\frac{1}{m}(a^\mu p^\nu-a^\nu p^\mu)
\end{equation}

Now, consider a polarized nucleon, e.g. proton, with momentum $p$
and polarization $a$, and define two quantities
--- the axial charge $g_A$ and the tensor charge $g_T$ in terms of average
values of quark axial and tensor currents similarly to
(\ref{ferm_av}), as follows:
\begin{equation} \label{axial_def}
\left\langle p,a \left| \hat{\bar u} \gamma_\mu\gamma^5 \hat u -
\hat{\bar d} \gamma_\mu\gamma^5 \hat d \right| p,a
\right\rangle=-g_A a_\mu,
\end{equation}
\begin{equation} \label{tensor_def}
\left\langle p,a \left| \hat{\bar u}\sigma_{\mu\nu}\gamma^5\hat u
- \hat{\bar d}\sigma_{\mu\nu}\gamma^5\hat d \right| p,a
\right\rangle =\frac{g_T}{m}(a^\mu p^\nu-a^\nu p^\mu)
\end{equation}
where $m$ is the proton mass.

In the limit of exact isotopic symmetry (which we assume further),
the isotopic structure of the axial current is $2\hat I_3$, and
its matrix element must coincide with that one of the current from
the same isotopic multiplet with structure $I_+$ taken between
proton and neutron. The latter transition arises in neutron
$\beta$--decay and we conclude that the constant $g_A$ must
coincide with the neutron decay axial constant $G_A$, which is
very well known experimentally\footnote{Our choice of the $G_A$
sign corresponds to \cite{Lee,Okun}} $G_A=1.27$. For the tensor
charge, unfortunately, we have no well established experimental
value, although it could be in principal measured as the first
momentum of the proton transversity distribution function $h_1(x)$
\cite{Ratcliffe}.

As for the theoretical calculations of the charges, it is well
known that the nonrelativistic quark model predicts
$g_A=\frac{5}{3}$, while for massless relativistic quarks in the
MIT bag model, one obtains much smaller values $g_A=1.09$, both in
disagreement with experiment. As for the tensor charge $g_T$,
various theoretical estimations give disparate results ranging
from 0.89 to 1.45, see \cite{Gamberg} for Refs. and discussions.
All this certainly stimulates further study of the question.

The approach proposed in this paper could be applied not only for
analysis of nucleon properties, but also for the calculation of
constants of hyperon semileptonic decays, which values are quite
important for the accurate determination of the CKM matrix
elements (see, e.g., \cite{Cabibbo}). We plan to develop a
systematic analysis of baryon properties in the framework of Dirac
orbital expansion in future papers; the first part of which is
reported below.

To construct the baryon wave function, one starts with the
Hamiltonian \cite{Simonov} obtained in the instantaneous
approximation from the general Bethe--Salpeter equation:
\begin{equation}
\Hat{H}\Psi(\rr_1,\rr_2,\rr_3)=E\Psi,\quad \Hat{H}=\sum_{i=1}^3
\Hat{H}_i+\Delta H \label{2}
\end{equation}
with
\begin{equation}
\Hat{H}_i=\mathbf{p}_{(i)}
\bsl{\alpha}_{(i)}+\beta_{(i)}(m_i+M(\rr_i)) \label{3}
\end{equation}
where $M(\rr_i)$ in the limit of vanishing gluon correlation
length is
\begin{equation}
M=\sigma |\rr_i-\rr_X|
\end{equation}
$\sigma$ is the string tension, and $\rr_X$ is the string-junction
coordinate. Here $\Delta H$ contains perturbative gluon exchanges.
We expand the baryon wave function in a series of products of
quark eigenfunctions $\psi_n^{(i)}=\binom{v^{(i)}}{w^{(i)}}$,
namely \cite{Weda}
\begin{equation}
\Psi(\rr_1,\rr_2,\rr_3)=\sum_{\{n_i\}}\prod_{i=1}^3
\psi_{n_i}^{(i)}(\rr_i)C_{n_1n_2n_3} \label{4}
\end{equation}

In what follows we shall consider the leading valence
approximation for the nucleon keeping only the first term in
(\ref{4}), $\Psi\to\Psi_0$, which contains the ground state
$S$--wave Dirac orbitals $|u_\lambda\rangle$ and
$|d_\lambda\rangle$ for $u$ and $d$ quarks with spins up and down.
So, for the proton polarized along the axis $Oz$, one has:
\begin{equation}
|p\uparrow\rangle=\sqrt{\frac{1}{18}}\left[-2(|u\uparrow u\uparrow
d\downarrow \rangle+\text{perm.})+(|u\uparrow u\downarrow
d\uparrow\rangle+\text{perm.} )\right] \label{5}
\end{equation}
This expression has the same form as  in the standard $SU(4)$ or
$SU(6)$ model \cite{Commins} except for the bispinor contents of
$|u_\lambda\rangle$ and $|d_\lambda\rangle$. In (\ref{5})
$u_\lambda$ and $d_\lambda$ orbitals depends on space coordinates
via the only function $\chi_\lambda$ in accordance with the
isotopic symmetry principle. Note, the proton state (\ref{5}) with
$S$--wave orbitals evidently has the average 3-momentum being
equal to zero:
\begin{equation}
\langle\mathbf{P}\rangle=\sum\langle\mathbf{p}_i\rangle=\mathbf{0}
\label{6}
\end{equation}
where $\mathbf{p}_i$ is the 3-momentum of the $i^{\text{th}}$
quark.

Insertion (\ref{5}) into (\ref{axial_def}) and (\ref{tensor_def})
yields\footnote{see \cite{Jaffe_Ji} for similar calculations in
bag models}:
\begin{equation}
g_A=\left\langle p\uparrow\left| \hat u^+\Sigma_3\hat u - \hat
d^+\Sigma_3\hat d \right| p\uparrow \right\rangle=
+\frac{4}{3}\langle \chi_\uparrow | \Sigma_3 |
\chi_\uparrow\rangle -\frac{1}{3}\langle \chi_\downarrow |
\Sigma_3 | \chi_\downarrow \rangle=+\frac{5}{3}\langle
\chi_\uparrow | \Sigma_3 | \chi_\uparrow\rangle \label{g_A_chi}
\end{equation}
\begin{equation}
g_T=\left\langle p\uparrow\left| \hat u^+\beta\Sigma_3\hat u -
\hat d^+\beta\Sigma_3\hat d \right| p\uparrow
\right\rangle=+\frac{4}{3}\langle \chi_\uparrow | \beta\Sigma_3 |
\chi_\uparrow\rangle -\frac{1}{3}\langle \chi_\downarrow |
\beta\Sigma_3 | \chi_\downarrow \rangle=+\frac{5}{3}\langle
\chi_\uparrow | \beta\Sigma_3 | \chi_\uparrow\rangle
\label{g_T_chi}
\end{equation}
where we use obvious equalities:
\begin{equation}
\langle \chi_\uparrow | \Sigma_3 | \chi_\uparrow\rangle=-\langle
\chi_\downarrow | \Sigma_3 | \chi_\downarrow \rangle,\qquad
\langle \chi_\uparrow | \beta\Sigma_3 |
\chi_\uparrow\rangle=-\langle \chi_\downarrow | \beta\Sigma_3 |
\chi_\downarrow \rangle
\end{equation}

To proceed further, we choose the standard representation of the
bispinor wave function
\begin{equation}
\Sigma_3=\begin{pmatrix} \sigma_3 & 0 \\ 0 & \sigma_3
\end{pmatrix},\qquad
\beta=\begin{pmatrix} 1 & 0 \\ 0 & -1 \end{pmatrix}
\end{equation}
\begin{equation}
\chi_\uparrow(r,\theta,\phi)=\frac{1}{r}
\binom{G(r)\Omega_{\frac{1}{2},0,\frac{1}{2}}(\theta,\phi)}
{F(r)\Omega_{\frac{1}{2},1,\frac{1}{2}}(\theta,\phi)} \label{chi}
\end{equation}
\begin{equation}
\Omega_{\frac{1}{2},0,\frac{1}{2}}(\theta,\phi)=
\binom{Y_{00}(\theta,\phi)}{0} ,\qquad
\Omega_{\frac{1}{2},1,\frac{1}{2}}(\theta,\phi)=
\binom{-\sqrt{\frac{1}{3}}Y_{10}(\theta,\phi)}{+\sqrt{\frac{2}{3}}Y_{11}(\theta,\phi)}
\label{Omega}
\end{equation}
The following normalization conditions hold:
\begin{equation}
\int \Omega_{j'l'm'}^+\Omega_{jlm}
do=\delta_{jj'}\delta_{ll'}\delta_{mm'},\qquad
\int\limits_0^\infty \left[G^2(r)+F^2(r)\right] dr=1
\end{equation}

To take into account perturbative gluon exchanges we represent
$\Delta H$ effectively as one-particle operators
\begin{equation}
\Delta H=\sum_{i=1}^3 \left(-\frac{\zeta}{r_i}\right) \label{8a}
\end{equation}
with an effective coupling constant $\zeta$ being fixed by
realistic $\alpha_s$ values (see below), and the equations for
$G(r)$, $F(r)$ acquire the form \cite{Popov}
\begin{equation}
\label{9}
\begin{gathered}
G'-\frac{1}{r}G-\left(\varepsilon+m+\sigma r+\frac{\zeta}{r}\right)F=0,\\
F'+\frac{1}{r}F+\left(\varepsilon-m-\sigma
r+\frac{\zeta}{r}\right)G=0
\end{gathered}
\end{equation}
where $m$ is the light quark mass and $\varepsilon$ is its energy.

Now, using equations (\ref{chi},\ref{Omega}), one can calculate
explicitly the matrix elements appeared in
(\ref{g_A_chi},\ref{g_T_chi}):
\begin{multline}
\langle \chi_\uparrow | \Sigma_3 | \chi_\uparrow\rangle=\\=\int
r^2drdo \left[\frac{G^2(r)}{r^2}\langle
\Omega_{\frac{1}{2},0,\frac{1}{2}} |\sigma_3|
\Omega_{\frac{1}{2},0,\frac{1}{2}} \rangle +
\frac{F^2(r)}{r^2}\langle \Omega_{\frac{1}{2},1,\frac{1}{2}}
|\sigma_3| \Omega_{\frac{1}{2},1,\frac{1}{2}} \rangle
\right]=\\=\int dr\left(G^2(r)-\frac{1}{3}F^2(r)\right)
\end{multline}
\begin{multline}
\langle \chi_\uparrow | \beta\Sigma_3 |
\chi_\uparrow\rangle=\\=\int r^2drdo
\left[\frac{G^2(r)}{r^2}\langle \Omega_{\frac{1}{2},0,\frac{1}{2}}
|\sigma_3| \Omega_{\frac{1}{2},0,\frac{1}{2}} \rangle -
\frac{F^2(r)}{r^2}\langle \Omega_{\frac{1}{2},1,\frac{1}{2}}
|\sigma_3| \Omega_{\frac{1}{2},1,\frac{1}{2}} \rangle
\right]=\\=\int dr\left(G^2(r)+\frac{1}{3}F^2(r)\right)
\end{multline}

Finally $g_A$ and $g_T$ can be written as\footnote{It is clear
from the aforesaid that neutron charges are simply opposite in
sign to the proton ones, so we do not need to consider them
separately}
\begin{equation}
g_A=+\frac{5}{3}\left(1-\frac{4}{3}\eta\right),\quad
g_T=+\frac{5}{3}\left(1-\frac{2}{3}\eta\right),\quad
\eta=\int\limits_0^\infty F^2(r) dr \label{10}
\end{equation}
Note, that in the nonrelativistic limit $\eta=0$ and
$g_A=g_T=5/3$.

\begin{table}
\begin{center}
\caption{\(g_A\), \(g_T\), and \(\eta\) for various theoretical
prescriptions in comparison with experimental data}
\begin{tabular}{|c|c|c|c|c|}
\hline
 & Exp. & NRQM &  \(\zeta=0\) &  \(\zeta=0.3\) \\
\hline \(g_A\) & 1.27 & 1.67 & 1.36 & 1.27 \\ \hline \(g_T\) & --
& 1.67 & 1.51 & 1.47 \\ \hline \(\eta\) & -- & 0 & 0.14 & 0.18 \\
\hline
\end{tabular}
\label{table_const}
\end{center}
\end{table}

The light quark mass $m$ is very small and can be omitted from
(\ref{9}) with good accuracy. So, the only dimensionful parameter
$\sigma$ remains in the model, but both $g_A$ and $g_T$ constants
are dimensionless and, correspondingly, do not depend on $\sigma$.
Thus, $\zeta$ proves to be the only essential model parameter. We
have computed the values of $\eta$, $g_A$, and $g_T$ for two
different values of $\zeta$: $\zeta=0$ and $\zeta=0.3$. The
results are given in Table \ref{table_const}; there also given for
comparison the values obtained in the nonrelativistic quark model.

One can see that the resulting $g_A$ is in the correct ballpark
for $\zeta\in [0,0.3]$. The value $\zeta=0.3$ corresponds to the
reasonable effective value of $\alpha_s$ in the $qq$ potential,
namely from
$\left\langle\sum\frac{2}{3}\frac{\alpha_s}{r_{ij}}\right\rangle=
\left\langle\sum\frac{\zeta}{r_i}\right\rangle$, and $\langle
r_{ij}\rangle\approx\sqrt{3}\langle r_i\rangle$, one has
\[
(\alpha_s)_{\text{eff}}=\frac{3\sqrt{3}}{4}\zeta\approx 0.39
\]
and this value of $\alpha_s$ was checked in the actual calculation
of the baryon masses in the Effective Hamiltonian approach
\cite{Narodetskii}. It is rewarding that the resulting $g_A=1.27$
is in close agreement with experiment.

Concerning the tensor charge $g_T$, one can compare these results
with the lattice data \cite{lat}, where both $g_A$ and $g_T$ are
close to each other and are in the interval $1.12\le g_A,g_T\le
1.18$ \cite{lat} for $m_\pi>0.5\text{~GeV}$.

There are two possible unaccounted effects which can influence our
results. First, the contribution of other terms in (\ref{4}) --
excited Dirac orbitals. The corresponding multichannel
calculations done in \cite{Weda} for magnetic moments, result in
decreasing of the modulus of magnetic moments of proton and
neutron by some $10-15\%$ when one accounts for four Dirac
orbitals for each quarks, and one can expect the same type of
corrections for $g_A$. Second, the contribution of chiral degrees
of freedom, i.e. of the $\pi$, $\eta$, $K$ exchanges. Again, for
nucleon magnetic moments these corrections are typically of the
order of $10\%$ \cite{Weda}, and we expect this to be an upper
limit for $g_A$, since magnetic moments are much more sensitive to
the contribution of the lowest Dirac components, than $g_A$ and
$g_T$, where these contributions enter quadratically and not
linearly.

These corrections are not taken into account above, which is
planned for a subsequent work, where also hyperon semileptonic
decays are considered \cite{in_prep}.

One should note, that relativistic approach to the baryon wave
function, based on the light-cone formalism \cite{Schlumpf} was
shown to improve the qualitative agreement of $g_A$ with
experiment, however quantitatively still far from experiment.

What is also possible to calculate in this approach are the baryon
decuplet masses. We restrict ourselves here only to the ground
state, so our leading $S$--wave approximation looks rather
reasonable. As the spin and the isospin of any pair of light
quarks in a decuplet baryon is equal to unity, one may expect the
pion exchange effects are suppressed for decuplet masses. Indeed,
calculations in \cite{Weda} show that pion exchanges yield a
negative shift to $m_\Delta$ by about 30~MeV. So, the baryon mass
is completely determined by the eigenvalues of the system of
equations (\ref{9}). The only modification is due to the existence
of a valence $s$-quark in a baryon -- in fact, we should calculate
independently the energy $\varepsilon$ and orbital $\chi$ for the
light quark $q$ and for the strange quark $s$; those are obtained
from the same equations by an obvious replacement: $m\to m_i$,
$\varepsilon\to\varepsilon_i$, $\chi\to\chi_i$, $i=q,s$ .

However, on this way there still exists a problem for the full
baryon wavefunction. We did not exclude clearly the baryon
center-of-mass motion from the wavefunction and provided only the
zero average value of the baryon 3-momentum, see (\ref{6}). This
means that the center-of-mass oscillates near zero point and bears
a nonzero kinetic energy, which must be excluded from the baryon
physical mass. The simplest way to do this is to define the baryon
mass in accordance with the Klein-Gordon equation:
\begin{equation}
M^2=E^2-\langle\mathbf{P}^2\rangle, \label{m1}
\end{equation}
where $\mathbf{P}=\sum\mathbf{p}_i$ is the total momentum of the
three-quark system, and $E=\sum\varepsilon_i$ is the total energy.

As we consider baryons in the leading valence approximation with
quarks in the $S$-wave orbitals, it is obvious that
\begin{equation}
\langle\mathbf{p}_i\mathbf{p}_j\rangle=0 \text{~for~} i\ne j,
\text{~and~}
\langle\mathbf{p}_i^2\rangle=\langle\chi_i|\mathbf{p}^2|\chi_i\rangle
\label{m2}
\end{equation}
where $\chi_i$ the eigenfunction of the system of equations
(\ref{9}). Thus
\begin{equation}
\hat H\chi_i=\left(\bsl{\alpha}\mathbf{p}+\beta\cdot m_i+\hat
1\cdot\left(-\frac{\zeta}{r}\right)+\beta\cdot\sigma
r\right)=\varepsilon_i\chi_i \label{m3}
\end{equation}
and the averaged momentum squared can be expressed via two simple
integrals:
\begin{multline}
\langle\chi_i|\mathbf{p}^2|\chi_i\rangle=
\langle\chi_i|\bsl{\alpha}\mathbf{p}|\bsl{\alpha}\mathbf{p}|\chi_i\rangle=
\left\langle\chi_i\left|\left(\varepsilon_i-\beta\cdot m_i-\hat
1\cdot\left(-\frac{\zeta}{r}\right)-\beta\cdot\sigma
r\right)^2\right|\chi_i\right\rangle=\\=
\left\langle\chi_i\left|\left(\varepsilon_i+\frac{\zeta}{r}\right)^2+\left(m_i+\sigma
r\right)^2 \right|\chi_i\right\rangle-
2\left\langle\chi_i\left|\left(\varepsilon_i+\frac{\zeta}{r}\right)\left(m_i+\sigma
r\right)\cdot\beta \right|\chi_i\right\rangle=\\= \int dr
\left(G^2(r)+F^2(r)\right)\left(\left(\varepsilon_i+\frac{\zeta}{r}\right)^2+\left(m_i+\sigma
r\right)^2\right)-\\-2\int dr \left(G^2(r)-F^2(r)\right)
\left(\varepsilon_i+\frac{\zeta}{r}\right)\left(m_i+\sigma
r\right) \label{m4}
\end{multline}

Now the baryon mass can be calculated as follows:
\begin{equation}
M=\sqrt{\left(\sum\varepsilon_i\right)^2-\sum\langle\mathbf{p}_i^2\rangle}
\label{m5}
\end{equation}
As before, we neglect here the light quark mass, so, to make the
numerical estimations, one needs to fix two dimensionful model
parameters: the string tension $\sigma$ and the strange quark mass
$m_s$. Being guided by our previous calculations of baryon
spectrum \cite{Narodetskii}, we use the following values:
\begin{equation}
\sigma=0.18~\text{GeV}^2,\quad m_s=210~\text{MeV}
\end{equation}

In the Table \ref{table_masses} we present the values of masses of
ground-state baryons, calculated in this approach in comparison
with experimental ones. One can see a fine agreement between the
data.
\begin{table}
\caption{Ground-state baryon masses, calculated in the Dirac
orbital model, in comparison with the experimental values (in
MeV)} \label{table_masses}
\begin{center}
\begin{tabular}{|c|c|c|c|c|}
\hline Baryon & $\Delta$ & $\Sigma^*$ & $\Xi^*$ & $\Omega$ \\
\hline Theor. & 1233 & 1381 & 1527 & 1672 \\ \hline Exp. & 1232 &
1383 & 1532 & 1672 \\ \hline
\end{tabular}
\end{center}
\end{table}

%\subsection*{Conclusions}

This work was supported in part by the Grant for support of
Leading Scientific Schools \# 843.2006.2, by the State Contract
\#~02.445.11.7424, and in part by the RFBR grants \#\#
05-02-17869, 06-02-17120, 06-02-17012.


\begin{thebibliography}{99}
\bibitem{Lee} T.D.Lee, \textit{Particle Physics and Introduction to Field
Theory} (Harwood Academic Publishers, 1981).
\bibitem{Okun} L.B.Okun, \textit{Leptons and Quarks} (North Holland,
1982).
\bibitem{Commins} E.D.Commins and P.H.Bucksbaum,
\textit{Weak Interactions of Leptons and Quarks} (Cambridge
University Press, 1983).
\bibitem{Bass} S.D.Bass, Eur. Phys. J. \textbf{A5}, 17 (1999).
\bibitem{Isgur} S.Capstick, N.Isgur, Phys. Rev. D \textbf{34},
2809 (1986).
\bibitem{Bonn} U.Loering, K.Kretzschmar, B.C.Metsch, H.R.Petry, Eur. Phys. J. \textbf{A10}, 309
(2001).
\bibitem{Ratcliffe} V.Barone, A.Drago, P.C.Ratcliffe, Phys. Rep.
\textbf{359}, 1 (2002).
\bibitem{Gamberg} L.Gamberg, G.Goldstein, Phys. Rev. Lett. \textbf{87},
242001 (2001); hep-ph/0106178.
\bibitem{Cabibbo} N.Cabibbo, E.C.Swallow, and R.Winston, hep-ph/0307214.
\bibitem{Simonov} Yu.A.Simonov, Phys. Atom. Nucl. \textbf{62}, 1932
(1999).
\bibitem{Weda} Yu.A.Simonov, J.A.Tjon, J.Weda, Phys. Rev. D
\textbf{65}, 094013 (2002); J.A.Tjon, J.Weda, Phys. Atom. Nucl.
\textbf{68}, 591 (2005).
\bibitem{Jaffe_Ji} R.L.Jaffe and X.--D.Ji, Nucl. Phys.
\textbf{B375}, 527 (1992).
\bibitem{Popov} V.D.Mur, V.S.Popov, Yu.A.Simonov, and V.P.Yurov,\\
Sov. Phys -- JETP \textbf{78}, 1 (1994).
\bibitem{Narodetskii} I.M.Narodetskii and M.A.Trusov, Phys. Atom.
Nucl. \textbf{67}, 762 (2004).
\bibitem{lat} A.Ali-Khan et al. [QCDSF-UKQCD Collab.],
Nucl. Phys. Proc. Suppl. \textbf{140}, 408 (2005);
hep-lat/0409161.
\bibitem{in_prep} Yu.A.Simonov and M.A.Trusov, in preparation.
\bibitem{Schlumpf} F.Schlumpf, Phys. Rev. D \textbf{51}, 2262 (1995).
\end{thebibliography}
\end{document}